\documentstyle[11pt,epsfig]{article}

\begin{document}

\begin{titlepage}
\begin{center}
\large Institute for Nuclear Research RAS, Moscow, Russia
\end{center}

\bigskip

\begin{flushright}
{ INR 0945a/98\\}
\end{flushright}     

\bigskip

\begin{center}{\Large 
Towards the Observation of Signal over Background
in Future Experiments}
\end{center}

\bigskip

  \begin{center}
{\large
    S.I.~Bityukov$^1$ (IHEP, Protvino RU-142284, Russia),\\
    N.V.~Krasnikov$^2$ (INR, Moscow 117312, Russia)
}
\end {center}

\vspace{2cm}
  
\begin{abstract}
We propose a method to estimate the probability of new physics
discovery in future high energy physics experiments. Physics
simulation gives both the average numbers $<N_b>$ of background 
and $<N_s>$ of signal events. We find that the proper definition
of the significance for $<N_b>,~<N_s>~\gg~1$ is
$S_{12} = \sqrt{<N_s>+<N_b>} - \sqrt{<N_b>}$ in comparison with often
used significances $S_1 = \displaystyle \frac{<N_s>}{\sqrt{<N_b>}}$ and
$S_2 = \displaystyle \frac{<N_s>}{\sqrt{<N_s> + <N_b>}}$.
We also propose a method for taking into account the systematical errors
related to nonexact knowledge of background and signal cross sections.
An account of such systematics is very essential in the search for 
supersymmetry at LHC.
\end{abstract}

\vspace{2cm}
\small
\noindent
\rule{3cm}{0.5pt}\\
$^1$E-mail: Serguei.Bitioukov\@cern.ch\\
$^2$E-mail: Nikolai.Krasnikov\@cern.ch\\
\vspace{1cm}

\begin{center}
     1998
\end{center}

\end{titlepage}

\newpage

\section{Introduction}

One of the common goals in the forthcoming experiments is the search for new 
phenomena. In the forthcoming high energy physics experiments 
(LHC, TEV22, NLC, ...) the main goal is the search for physics
beyond the Standard Model (supersymmetry, $Z'$-, $W'$-bosons, ...) and
the Higgs boson discovery as a final confirmation of the Standard Model.
In estimation of the discovery potential of the future experiments
(to be specific in this paper we shall use as an example CMS experiment
at LHC~\cite{1}) the background cross section 
is calculated and for the given integrated luminosity $L$ the average 
number of background events is $<N_b> = \sigma_b \cdot L$.
Suppose the existence of a new physics 
leads to the nonzero signal cross section $\sigma_s$ with the same 
signature as for the background cross section  that 
results in  the prediction
of the additional average number of signal events 
$<N_s> = \sigma_s \cdot L$ for the integrated luminosity $L$.

The total average number of the events is 
$<N_{ev}> = <N_s> + <N_b> = (\sigma_s + \sigma_b) \cdot L$.
So, as a result of new physics existence, we expect an excess 
of the average number of events. In real experiments the probability
of the realization of $n$ events is described by Poisson 
distribution~\cite{2}

\begin{equation}
f(n,<n>) = \frac{<n>^n}{n!} e^{-<n>}.
\end{equation}

Here $<n>$ is the average number of events.

Remember that the Poisson distribution $f(n,<n>)$ gives~\cite{3} the 
probability of finding exactly $n$ events in the given interval
of (e.g. space and time) when the events occur independently
of one another and of $x$ at an average rate  of $<n>$ per the 
given interval. For the Poisson distribution the variance $\sigma^2$
equals to $<n>$. So, to estimate the probability of the new physics discovery 
we have to compare the Poisson statistics with $<n> = <N_b>$ and
$<n> = <N_b> + <N_s>$. Usually, high energy physicists use the following  
``significances'' for testing the possibility to discover new physics
in an  experiment:

\begin{itemize}
\item[(a)]
``significance'' $S_1 = \displaystyle \frac{N_s}{\sqrt{N_b}}$~\cite{4},
\item[(b)]
``significance'' $S_2 = \displaystyle \frac{N_s}{\sqrt{N_s + N_b}}$~\cite{5,6}.
\end{itemize}

A conventional claim is that for $S_1~(S_2) \ge 5$ we shall discover
new physics (here, of course, the systematical errors are ignored).
For $N_b \gg N_s$ the significances  $S_1$ and $S_2$ coincide (the 
search for Higgs boson through the $H \rightarrow \gamma \gamma$
signature). For the case when $N_s \sim N_b$,~$S_1$ and $S_2$ differ.
Therefore, a natural question arises: what is the correct definition for the 
significance $S_1$, $S_2$ or anything else ?

It should be noted  that there is a crucial difference between ``future''
experiment and the ``real'' experiment. In the ``real'' experiment the total 
number of events $N_{ev}$ is a given number (already has been measured)
and we compare it with $<N_b>$ when we test the validity of the
standard physics. So, the number of possible signal events is determined
as $N_s = N_{ev} - <N_b>$ and it is compared  with the average number of 
background events $<N_b>$. The fluctuation of the background is 
$\sigma_{fb} = \sqrt{N_b}$, therefore, we come to the $S_1$ significance
as the measure of the distinction from the standard physics.
In the conditions of the ``future'' experiment when we want to search
for new physics, we know only the average number of the background
events and the average number of the signal events, so we
have to compare the Poisson distributions $P(n,<N_b>)$ and
$P(n,<N_b>+<N_s>)$ to determine the probability to find new physics
in the future experiment.

In this paper we estimate the probability to discover new physics
in future experiments. We show that for $<N_s>,~<N_b> \gg 1$ the
proper determination of the significance is 
$S = \sqrt{<N_b>+<N_b>} - \sqrt{<N_b>}$. We also suggest a method
which takes into account systematic errors related to  nonexact
knowledge of the signal and background cross sections. 

The organization 
of the paper is the following. In the next section we give a method
for the determination of the probability to find new physics in the future
experiment and calculate the probability to discover new physics
for the given $(<N_b>,~<N_s>)$ numbers of background and signal events
under the assumption that there are no systematic errors. In section 3 
we estimate the influence of the systematics related to the nonexact 
knowledge of the signal and background cross sections on the probability 
to discover new physics in future experiments. Section 4 contains the 
concluding remarks.

\section{An analysis of statistical fluctuations}

Suppose that for some future experiment we know the average number of 
the background and signal event $<N_b>,~<N_s>$. As it has been mentioned
in the Introduction, the probability of realization of $n$ events in an 
experiment is given by the Poisson distribution

\begin{equation}
P(n,<n>) = \frac{<n>^n}{n!} e^{-<n>},
\end{equation}
where $<n> = <N_b>$ for the case of the absence of new physics
and $<n> = <N_b> + <N_s>$ for the case when new physics exists.
So, to determine the probability to discover new physics in future
experiment, we have to compare the Poisson distributions with
$<n> = <N_b>$ (standard physics) and $<n> = <N_b> + <N_s>$ (new physics).

Consider, at first, the case when $<N_b>~~\gg 1$, $<N_s>~~\gg 1$.
In this case the Poisson distributions approach the Gaussian distributions

\begin{equation}
P_G(n,\mu,\sigma^2) = \frac{1}{\sigma \sqrt{2 \pi}} \cdot
e^{-\frac{(n - \mu)^2}{2 \sigma^2}},
\end{equation}

with $\mu = \sigma^2$ and $\mu = <N_b>$ or $\mu = <N_b> + <N_s>$. 
Here $n$ is a real number.

The Gaussian distribution describes  the probability density to realize
$n$ events in the future experiment provided the average number of events
$<n>$ is a given number. In Fig.1 we show two Gaussian distributions 
$P_G$ with $<n> = <N_b> = 53$ and $<n> = <N_b> + <N_s>$ = 104
(\cite{6}, Table.13, cut 6). As is clear from Fig.1 
the common area for these two curves (the first curve shows the
``standard physics'' events distribution and the second one 
gives the ``new physics'' events distribution) is the probability 
that ``new physics'' can be described by the ``standard physics''.
In other words, suppose we know for sure that new physics takes place
and the probability density of the events realization is described
by curve II ($f_2(x)$). The probability $\kappa$ that the ``standard physics''
(curve I ($f_1(x)$)) can imitate  new physics 
(i.e. the probability that we measure ``new physics''
but we think that it is described by the ``standard physics'')
is described by common area of curve I and II. 

\begin{figure}[ht]
\epsfig{file=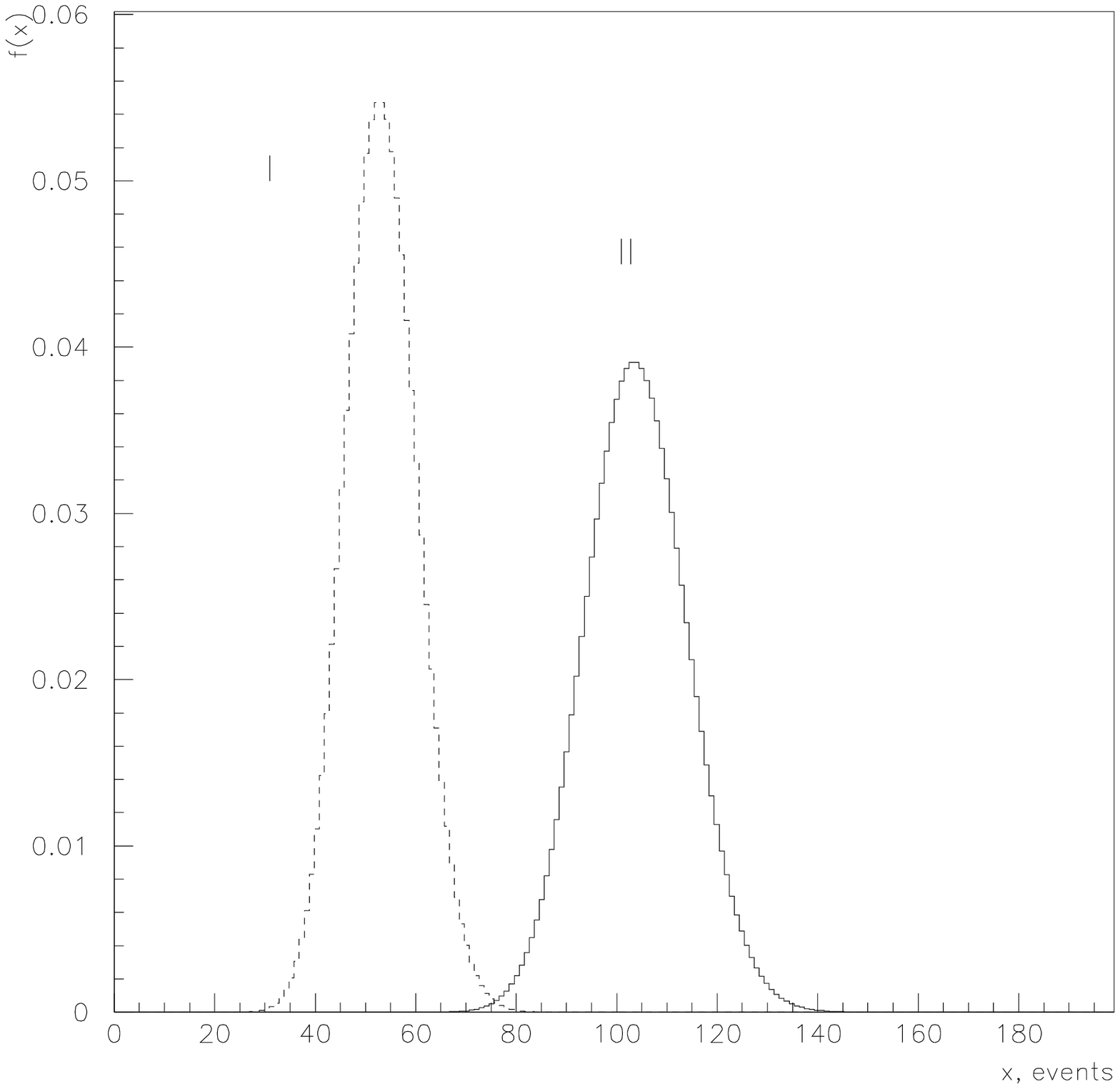,width=\textwidth}
\caption{The probability density functions
$f_{1,2}(x) \equiv P_G(x,\mu_{1,2},\sigma^2)$ for $\mu_1 = <N_b> = 53$ 
and $\mu_2 = <N_b> + <N_s> = 104$.}
\label{fig.1}
\end{figure}
        
Numerically, we find that 

\begin{eqnarray}
\kappa &=& \frac{1}{\sqrt{2 \pi} \sigma_2}
\int_{-\infty}^{\nonumber \sigma_1 \sigma_2}exp[-\frac{(x-\sigma_2^2)^2}{2 \sigma_2^2}]dx +
\frac{1}{\sqrt{2 \pi} \sigma_1}
\int_{\sigma_1 \sigma_2}^{\infty}exp[-\frac{(x-\sigma_1^2)^2}{2 \sigma_1^2}]dx\\  
&=&\frac{1}{\sqrt{2 \pi}}
[\int_{-\infty}^{\sigma_1-\sigma_2}exp[-\frac{y^2}{2}]dy +
\int_{\sigma_2-\sigma_1}^{\infty}exp[-\frac{y^2}{2}]dy] \\ \nonumber
&=&1 - erf(\frac{\sigma_2-\sigma_1}{\sqrt{2}}).
\end{eqnarray}

Here $\sigma_1 = \sqrt{N_b}$ and $\sigma_2 = \sqrt{N_b + N_s}$.

As follows from formula (4) the role of the significance $S$
plays 

\begin{equation}
S_{12} = \sigma_2 - \sigma_1 = \sqrt{N_b + N_s} - \sqrt{N_b}. 
\end{equation}

Note that  in refs.\cite{7}  the following criterion of the signal
discovery has been used. The signal was assumed to be observable
if $(1 - \epsilon) \cdot 100\%$ upper confidence level for the
background event rate is equal to $(1 - \epsilon) \cdot 100\%$
lower confidence level for background plus signal
$(\epsilon = 0.01 - 0.05)$. The corresponding
significance is similar to our significance $S_{12}$. The
difference is that in our approach the probability
density $\kappa$ that new physics is described by standard physics is
equal to $2 \epsilon$. 

It means that for 
$S_{12} = 1, 2, 3, 4, 5, 6$ the probability $\kappa$ is correspondingly
$\kappa = 0.31, 0.041, 0.0027, 6.3 \cdot 10~^{-5}, 5.7 \cdot~10^{-7},
2.0 \cdot~10^{-7}$ in accordance with a general picture. 
As it has been mentioned in the Introduction two definitions of the significance
are mainly used in the literature: 
 $S_1 = \displaystyle \frac{N_s}{\sqrt{N_b}}$~\cite{4} and
 $S_2 = \displaystyle \frac{N_s}{\sqrt{N_s + N_b}}$~\cite{5}.
The significance $S_{12}$ is expressed in terms of the significances
$S_1$ and $S_2$ as $S_{12} = \displaystyle \frac{S_1 S_2}{S_1 + S_2}$.

For $N_b \gg N_s$ (the search for Higgs boson through 
$H \rightarrow \gamma \gamma$ decay mode) we find that 

\begin{equation}
S_{12} \approx 0.5~S_1 \approx 0.5~S_2.
\end{equation}

It means that for $S_1 = 5$ (according to a common convention the
$5 \sigma$ confidence level means a new physics discovery) the
real significance is $S_{12} = 2.5$, that  corresponds to $\kappa = 1.2 \%$.

For the case $N_s = k N_b$, $S_{12} = k_{12} S_2$, where for
$k = 0.5, 1, 4, 10$ the value of $k_{12}$ is 
$k_{12} = 0.55, 0.59, 0.69, 0.77$. For not too high values of 
$<N_b>$ and $<N_b + N_s>$, we have to compare the Poisson distributions
directly. Again for the Poisson distribution $P(n,<n>)$ with the area
of definition for nonnegative integers we can define
$P(x,<n>)$ for real $x$ as 

\begin{equation}
\displaystyle \tilde P(x, <n>) = \left\{
\begin{array}{ll}
               0,        & x \le 0, \\
     P([x], <n>),        & x > 0.    
\end{array}
\right.
\end{equation}

It is evident that 

\begin{equation}
\int_{-\infty}^{\infty}{\tilde P(x, <n>) dx} = 1.
\end{equation}

So, the generalization of the previous determination of $\kappa$
in our case is straightforward, namely, $\kappa$ is nothing but 
the common area of the curves described by $\tilde P(x, <N_b>)$
(curve I) and $\tilde P(x, <N_b> + <N_s>)$ (curve II) (see, Fig.2).

\begin{figure}[ht]
\epsfig{file=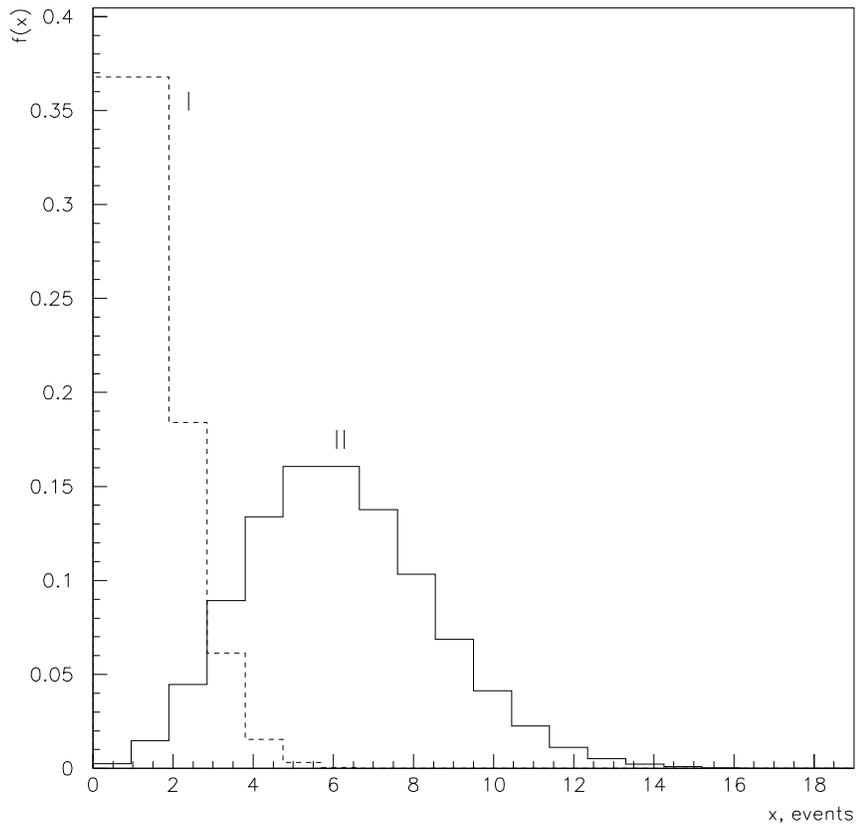,width=\textwidth}
\caption{The probability density functions
$f_{1,2}(x) \equiv \tilde P(x,\mu_{1,2})$ for $\mu_1$ = $<N_b>$ = $1$ 
and $\mu_2 = <N_b> + <N_s> = 6$.}
\label{fig.2}
\end{figure}

One can find that \footnote{We are indented to Igor Semeniouk for the help 
in the derivation of these formulae}

$ \kappa = \kappa_{1} + \kappa_{2}, $

$\displaystyle \kappa_{1}=  \sum_{n = n_0 +1}^{\infty}\frac{(<N_b>)^n}{n!}
\displaystyle e^{-<N_b>} = 
1 - \frac{\Gamma(n_0+1, <N_b>)}{\Gamma(n_0+1)} $,

$\displaystyle \kappa_2 = \sum^{n_0}_{n=0}\frac{(<N_b> + <N_s>)^n}{n!}
\displaystyle e^{-(<N_b> + <N_s>)}, $


$\displaystyle n_0 = [ \frac{<N_s>}{ln(1 + \frac{<N_s>}{<N_b>})}].$

Numerical results  are presented  in Tables 1-6. 

As it follows
from these Tables for finite values of $<N_s>$ and $<N_b>$
the deviation from asymptotic formula (4) is essential.
For instance, for $N_s = 5$, $N_b = 1~~(S_1 = 5)$~
$\kappa = 14\%$. For $N_s = N_b = 25~~(S_1 = 5)$~
$\kappa = 3.9\%$, whereas asymptotically for $N_s \gg 1$ we find 
$\kappa = 1.2\%$. 
Similar situation takes place for $N_s \sim N_b$.

\section{An account of systematic errors related to nonexact
knowledge of background and signal cross sections}

In the previous section we determined the statistical error $\kappa$
(the probability that ``new physics'' is described by ``standard physics'').
In this section we investigate the influence of the systematical errors
related to a nonexact knowledge of the background and signal 
cross sections on the probability $\kappa$ not to confuse a new
physics with the old one.

Denote the Born background and signal cross sections as 
$\sigma_b^0$ and $\sigma_s^0$. An account of one loop corrections leads to
$\sigma_b^0 \rightarrow \sigma_b^0(1+\delta_{1b})$ and   
$\sigma_s^0 \rightarrow \sigma_s^0(1+\delta_{1s})$, where typically 
$\delta_{1b}$ and $\delta_{1s}$ are {\cal O}(0.5).

Two loop corrections at present are not known. So, we can assume that the 
uncertainty related with nonexact knowledge of cross sections is around
$\delta_{1b}$ and $\delta_{1s}$ correspondingly. In other words, we assume that
the exact cross sections lie in the intervals 
$(\sigma_b^0, \sigma_b^0 (1+2 \delta_{1b}))$ and  
$(\sigma_s^0, \sigma_s^0 (1+2 \delta_{1s}))$.
The average number of background and signal events lie in the intervals  

\begin{equation}
(<N_b^0>, <N_b^0> (1+2 \delta_{1b})) 
\end{equation}

and  

\begin{equation}
(<N_s^0>, <N_s^0> (1+2 \delta_{1s})), 
\end{equation}

where $<N_b^0> = \sigma_b^0 \cdot L$, $<N_s^0> = \sigma_s^0 \cdot L$. 

To determine the probability that the new physics is described by the 
old one, we 
again have to compare  two Poisson distributions with and without new physics
but in distinction from Section 2 we have to compare the Poisson distributions 
in which the average numbers lie in some intervals. So, a priori the only
thing we know is that the average numbers of background and signal events
lie in the intervals (9) and (10), but we do not know the exact values of $<N_b>$
and $<N_s>$. To determine the probability that 
the new physics is described by the old,
consider the worst case when we think that new  physics 
is described by the minimal number of average events 

\begin{equation}
<N_b^{min}> = <N_b^0> + <N_s^0>.
\end{equation}

Due to the fact that we do not know the exact value of the background
cross section,  consider the worst case when the average 
number of background events is equal to $<N_b^0> (1 + 2 \delta_{1b})$.
So, we have to compare the Poisson distributions with 
$<n> = <N_b^0> + <N_s^0> =$

$<N_b^0> (1 + 2 \delta_{1b}) + (<N_s^0>  - 2 \delta_{1b} <N_b^0>)$
and $<n> = <N_b^0> (1 + 2 \delta_{1b})$.
Using the result of the previous Section, we find that for case  
$<N_b^0>~\gg~1,~<N_s^0>~\gg~1$  the effective significance is

\begin{equation}
S_{12s} = \sqrt{<N_b^0>+<N_s^0>} - \sqrt{<N_b^0>(1+2\delta_{1b})}.
\end{equation}

For the limiting case $\delta_{1b} \rightarrow 0$, we reproduce
formula (5). For not too high values of $<N_b^0>$ and $<N_s^0>$, we
have to use the results of the previous section (Tables 1-6).

As an example consider the case when $\delta_{1b} = 0.5$,
$<N_s> = 100$, $<N_b> = 50$ (typical situation for sleptons search). 
In this case we find that\\

 $S_1 = \displaystyle \frac{<N_s>}{\sqrt{<N_b>}} = 14.1,$

 $S_2 = \displaystyle \frac{<N_s>}{\sqrt{<N_s> + <N_b>}} = 8.2$

 $S_{12} = \sqrt{<N_b>+<N_s>} - \sqrt{<N_b>} = 5.2,$

 $S_{12s} = \sqrt{<N_b>+<N_s>} - \sqrt{2<N_b>} = 2.25.$\\

The difference between CMS adopted significance
$S_2~=~8.2$ (that corresponds to the probability 
$\kappa = 0.206 \cdot 10^{-6}$)
and the significance $S_{12s}~=~2.25$ taking into account 
systematics related to  nonexact knowledge of background cross section 
is factor 3.6 
The direct comparison of the Poisson distributions 
with $<N_b>(1 + 2\delta_{1b}) = 100$ and $<N_b>(1 + 2\delta_{1b})
 + <N_{s,eff}>$ ( $<N_{s,eff}> = <N_s> - 2\delta_{1b}<N_b> = 50)$ 
gives $\kappa_{s} 
= 0.0245$.

Another example is with $<N_s> = 28$, $<N_b>=8$ and $\delta_{1b} = 0.5$. 
For such example we have $S_1 = 9.9$, $S_2 = 4.7$, $S_{12} = 3.2$, 
$S_{12s} = 2.0$, $\kappa_{s} =0.045$.

So, we see that an account of the systematics related to nonexact knowledge 
of background cross sections is very essential and it decreases the LHC SUSY 
discovery potential.

\section{Conclusions}

In this paper we determined the probability to discover the new physics 
in the future experiments when the average number of background $<N_b>$ and 
signal events  $<N_s>$ is known. We have found that in this case for 
$<N_s>~\gg~1$ and $<N_b>~\gg~1$ the role of significance plays 

 $S_{12}~=~\sqrt{<N_b>+<N_s>}~-~\sqrt{<N_b>}$ 

\noindent
in comparison with often used 
expressions for the significances
 $S_1~=~\displaystyle~\frac{<N_s>}{\sqrt{<N_b>}}$ and
 $S_2~=~\displaystyle~\frac{<N_s>}{\sqrt{<N_s>~+~<N_b>}}$.

For $<N_s>~\ll~<N_b>$ we have found that $S_{12}  = 0.5 S_1 = 0.5 S_2$.
For not too high values of $<N_s>$ and $<N_b>$, when the deviations
from the Gaussian distributions are essential, our results are presented
in Tables 1-6. We also proposed a method for taking into account
systematical errors related to the nonexact knowledge of background and signal
events. An account of such kind of systematics is very essential in the search 
for supersymmetry and leads to an essential decrease in the probability to 
discover the new physics in the future experiments.

We are indebted to M.Dittmar for very hot discussions and useful questions
which were one of the motivations to perform this study. We are grateful
to V.A.Matveev for the interest and useful comments.

\begin{table}[h]
\small
    \caption{ The dependence of $\kappa$ on $<N_s>$ and $<N_b>$ for 
$S_1 =5$} 
    \label{tab:Tab.1}
    \begin{center}
\begin{tabular}{|r|r|r|}
\hline
$<N_s>$ & $<N_b>$ & $\kappa$\\ 
\hline
   5 &  1   &  $0.1420$ \\
  10 &  4   &  $0.0828$ \\
  15 &  9   &  $0.0564$ \\
  20 & 16   &  $0.0448$ \\
  25 & 25   &  $0.0383$ \\
  30 & 36   &  $0.0333$ \\
  35 & 49   &  $0.0303$ \\
  40 & 64   &  $0.0278$ \\
  45 & 81   &  $0.0260$ \\
  50 & 100  &  $0.0245$ \\
  55 & 121  &  $0.0234$ \\
  60 & 144  &  $0.0224$ \\
  65 & 169  &  $0.0216$ \\
  70 & 196  &  $0.0209$ \\
  75 & 225  &  $0.0203$ \\
  80 & 256  &  $0.0198$ \\
  85 & 289  &  $0.0193$ \\
  90 & 324  &  $0.0189$ \\
  95 & 361  &  $0.0185$ \\
 100 & 400  &  $0.0182$ \\
 150 & 900  &  $0.0162$ \\
 500 & $10^4$ &  $0.0136$ \\
 5000 & $10^6$ &  $0.0125$ \\
\hline
\end{tabular}
    \end{center}
\end{table}

\begin{table}[h]
\small
    \caption{ The dependence of $\kappa$ on $<N_s>$ and $<N_b>$ 
for $S_2 \approx 5$.}
    \label{tab:Tab.2}
    \begin{center}
\begin{tabular}{|l|l|l|}
\hline
$<N_s>$ & $<N_b>$ & $\kappa$ \\ 
\hline
  26 &   1 &    $0.154 \cdot 10^{-4}$ \\
  29 &   4 &    $0.142 \cdot 10^{-3}$ \\
  33 &   9 &    $0.440 \cdot 10^{-3}$ \\
  37 &  16 &    $0.993 \cdot 10^{-3}$ \\
  41 &  25 &    $0.172 \cdot 10^{-2}$ \\
  45 &  36 &    $0.262 \cdot 10^{-2}$ \\
  50 &  49 &    $0.314 \cdot 10^{-2}$ \\
  55 &  64 &    $0.357 \cdot 10^{-2}$ \\
 100 & 300 &    $0.735 \cdot 10^{-2}$ \\
 150 & 750 &    $0.894 \cdot 10^{-2}$ \\
\hline
\end{tabular}
    \end{center}
\end{table}

\begin{table}[h]
\small
    \caption{$<N_s> = \displaystyle {1 \over 5} \cdot <N_b>$.
The dependence of $\kappa$ on $<N_s>$ and $<N_b>$.}
    \label{tab:Tab.3}
    \begin{center}
\begin{tabular}{|l|l|l|l|l|l|}
\hline
$<N_s>$ & $<N_b>$ & $\kappa$\\ 
\hline
   50&  250&   $0.131$ \\
  100&  500&   $0.033$ \\
  150&  750&   $0.83 \cdot 10^{-2}$ \\
  200& 1000&   $0.24 \cdot 10^{-2}$ \\
  250& 1250&   $0.61 \cdot 10^{-3}$ \\
  300& 1500&   $0.22 \cdot 10^{-3}$ \\
  350& 1750&   $0.50 \cdot 10^{-4}$ \\
  400& 2000&   $0.10 \cdot 10^{-4}$ \\
\hline
\end{tabular}
    \end{center}
\end{table}

\begin{table}[h]
\small
    \caption{$<N_s> = \displaystyle {1 \over 10} \cdot <N_b>$. 
The dependence of $\kappa$ on $<N_s>$ and $<N_b>$.}
    \label{tab:Tab.4}
    \begin{center}
\begin{tabular}{|l|l|l|}
\hline
$<N_s>$ & $<N_b>$ &  $\kappa$\\ 
\hline
   50&  500&   $0.274$ \\
  100& 1000&   $0.123$ \\
  150& 1500&   $0.057$ \\
  200& 2000&   $0.029$ \\
  250& 2500&   $0.014$ \\
  300& 3000&   $0.75 \cdot 10^{-2}$ \\
  350& 3500&   $0.36 \cdot 10^{-2}$ \\
  400& 4000&   $0.20 \cdot 10^{-2}$ \\
  450& 4500&   $0.10 \cdot 10^{-2}$ \\
  500& 5000&   $0.50 \cdot 10^{-3}$ \\
\hline
\end{tabular}
    \end{center}
\end{table}

\begin{table}[h]
\small
    \caption{$<N_s> = <N_b>$. 
The dependence of $\kappa$ on $<N_s>$ and $<N_b>$.}
    \label{tab:Tab.5}
    \begin{center}
\begin{tabular}{|l|l|l|}
\hline
$<N_s>$ & $<N_b>$ &  $\kappa$\\ 
\hline
    2. &   2. &     0.562    \\
    4. &   4. &     0.406   \\
    6. &   6. &     0.308   \\
    8. &   8. &     0.241    \\
   10. &  10. &     0.187  \\
   12. &  12. &     0.150  \\
   14. &  14. &     0.119    \\
   16. &  16. &     0.098 \\
   18. &  18. &     0.079 \\
   20. &  20. &     0.064 \\
   24. &  24. &     0.043 \\
   28. &  28. &     0.027 \\
   32. &  32. &     0.018 \\
   36. &  36. &     0.014 \\
   40. &  40. &    $0.84 \cdot 10^{-2}$ \\
   50. &  50. &    $0.33 \cdot 10^{-2}$ \\
   60. &  60. &    $0.13 \cdot 10^{-2}$ \\
   70. &  70. &    $0.47 \cdot 10^{-3}$ \\
   80. &  80. &    $0.16 \cdot 10^{-3}$ \\
  100. & 100. &    $0.30 \cdot 10^{-4}$ \\
\hline
\end{tabular}
    \end{center}
\end{table}

\begin{table}[h]
\small
    \caption{$<N_s> = 2 \cdot <N_b>$. 
The dependence of $\kappa$ on $<N_s>$ and $<N_b>$.}
    \label{tab:Tab.6}
    \begin{center}
\begin{tabular}{|l|l|l|}
\hline
$<N_s>$ & $<N_b>$ &  $\kappa$\\ 
\hline
    2. &   1. &     0.464    \\
    4. &   2. &     0.295   \\
    6. &   3. &     0.199   \\
    8. &   4. &     0.143    \\
   10. &   5. &     0.101  \\
   12. &   6. &     0.074  \\
   14. &   7. &     0.050    \\
   16. &   8. &     0.038 \\
   18. &   9. &     0.027 \\
   20. &  10. &     0.020 \\
   24. &  12. &     0.011 \\
   28. &  14. &    $0.60 \cdot 10^{-2}$ \\
   32. &  16. &    $0.35 \cdot 10^{-2}$ \\
   36. &  18. &    $0.19 \cdot 10^{-2}$ \\
   40. &  20. &    $0.85 \cdot 10^{-3}$ \\
   50. &  25. &    $0.27 \cdot 10^{-3}$ \\
   60. &  30. &    $0.40 \cdot 10^{-4}$ \\
\hline
\end{tabular}
    \end{center}
\end{table}

\end{document}